# BlueScript: A Disaggregated Virtual Machine for Microcontrollers

Fumika Mochizuki[a] , Tetsuro Yamazaki[a] , and Shigeru Chiba[a]

a   The University of Tokyo, Tokyo, Japan

**Abstract**    Virtual machines (VMs) are highly beneficial for microcontroller development. In particular, interactive programming environments greatly facilitate iterative development processes, and higher execution speeds expand the range of applications that can be developed. However, due to their limited memory size, microcontroller VMs provide a limited set of features. Widely used VMs for microcontrollers often lack interactive responsiveness and/or high execution speed. While researchers have investigated offloading certain VM components to other machines, the types of components that can be offloaded are still restricted.

In this paper, we propose a disaggregated VM that offloads as many components as possible to a host machine. This makes it possible to exploit the abundant memory of the host machine and its powerful processing capability to provide rich features through the VM. As an instance of a disaggregated VM, we design and implement a BlueScript VM. The BlueScript VM is a virtual machine for microcontrollers that provides an interactive development environment. We offload most of the components of the BlueScript VM to a host machine. To reduce communication overhead between the host machine and the microcontroller, we employed a data structure called a shadow machine on the host machine, which mirrors the execution state of the microcontroller.

Through our experiments, we confirmed that offloading components does not seriously compromise their expected benefits. We assess that an offloaded incremental compiler results in faster execution speed than MicroPython and Espruino, while keeping interactivity comparable with MicroPython. In addition, our experiments observe that the offloaded dynamic compiler improves VM performance. Through this investigation, we demonstrate the feasibility of providing rich features even on VMs for memory-limited microcontrollers.

## The Art, Science, and Engineering of Programming

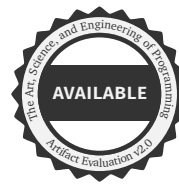 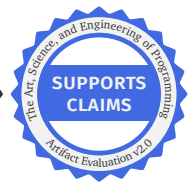

 **Introduction**

Virtual machines (VMs) have seen remarkable recent advancements, resulting in the incorporation of a wide variety of features. For example, V8,[1] an instance of the JavaScript VM, provides various features, including high execution speed, automatic memory management, run-time type checking, exception handling, asynchronous processing, interactive programming, and portability. These features are significantly beneficial for programmers. They enable rapid development of high-performance and secure programs. In particular, interactive programming is useful for a wide range of software development. It assists rapid prototyping based on trial and error, for example, parameter adjustments for rendering a web page or controlling peripheral devices.

In contrast, the VM design for microcontrollers is limited by the amount of available memory and processing power, restricting the features provided by the VM. For example, MicroPython [5], an instance of the Python VM for microcontrollers, and Espruino [22], an instance of the JavaScript VM for microcontrollers, provide interactive programming and other beneficial features, but they demonstrate significantly lower performance compared to the C language because of a lack of advanced compilation features such as dynamic compilation. The C programming environment, which is popular for microcontroller programming, can be considered as a significant streamlining of virtual machine because program execution requires only a small run-time library regarded as a VM on a microcontroller after cross-platform compilation and linking on a development machine. Although this VM achieves high execution speed, it does not provide a number of beneficial features such as interactive programming, automatic memory management, and run-time type checking.

Although offloading VM components to more resource-rich machines has been proposed to address this problem, only a limited subset of components has been offloaded to date. This approach involves offloading certain VM components to servers or host machines (programmers' laptop PCs) with substantially more resources. Although offloading of just-in-time compilers and debuggers has been suggested, numerous components still reside on the microcontroller, indicating considerable potential for the development of more comprehensive offloading techniques.

This paper proposes a disaggregated VM, which offloads as many components as possible to a host machine. The disaggregated VM exploits the abundant resources of the host machine to provide rich functionality. As an instance of this disaggregated VM, we design and implement the BlueScript VM. BlueScript is our own language similar to TypeScript. The BlueScript VM is designed for microcontrollers, and it offers an interactive programming environment. A BlueScript program is executed interactively through a Read-Eval-Print Loop (REPL) with an incremental compiler and a dynamic compiler. The VM components retained on the microcontroller are only a memory manager, a profiler, an execution kernel, and a hardware library for wireless network communication. The BlueScript VM offloads the rest of the VM components, including the incremental compiler and the dynamic compiler. A key component of the BlueScript VM implementation is *a shadow machine*. It is a data structure on the host machine and reflects the current execution state of the microcontroller. The information provided by the shadow machine is used to execute the other components on the host machine, and it mitigates the impact on the communication overhead between the microcontroller and the host machine. The abstract syntax tree of a running program and its symbol table used by a linker are examples of the information reflected by the shadow machine. It is not a simple cached copy of the memory image on the microcontroller. Our experiments show that the incremental compiler and dynamic compiler were successfully offloaded without significantly compromising their benefits, despite the potential overhead of offloading.

---
[1] https://v8.dev (Visited on 2025-09-29).

This paper makes the following contributions.

- We propose a new virtual machine (VM) architecture, called a disaggregated VM, designed specifically for microcontrollers. This VM leverages the abundant resources of a host machine to provide beneficial features for resource-constrained microcontrollers.

- We present the design and implementation of the BlueScript VM as a concrete instance of a disaggregated VM. This implementation introduces a component called *the shadow machine*, which helps reduce communication overhead between the host machine and the microcontroller.

- We demonstrate that the offloaded incremental compiler results in faster execution speed than MicroPython and Espruino, while keeping interactivity comparable with MicroPython. In addition, our experiments observe that the offloaded dynamic compiler improves VM performance.

## 2  Virtual Machines for Microcontrollers

Due to the limited memory of microcontrollers, it is essential to keep a language virtual machine (VM) for microcontrollers as small as possible. Microcontrollers have constrained memory. For instance, the ESP32 microcontroller from Espressif Systems has 520KB of SRAM, and the RP2040 microcontroller used for Raspberry Pi Pico has only 264KB. The ESP32 has 16MB of flash memory, and the Raspberry Pi Pico has a mere 2MB. Furthermore, applications running within the VM also necessitate memory space. A large VM footprint reduces the memory space available for applications, thereby limiting the complexity and scope of applications that programmers can develop.

A common approach for minimizing the VM size is to reduce its functionality. MicroPython [5], for example, sacrifices execution speed because of a lack of a dynamic compiler or a Just-In-Time (JIT) compiler. In our experiments using the "Are We Fast Yet?" [15] benchmark suite, MicroPython exhibits an average execution time approximately 100 times longer than equivalent C code. Programming a microcontroller in the C language is typically a cross-platform development, where the run-time execution environment can be viewed as a minimalistic VM that provides only the most essential functionalities. This environment provides the highest possible execution performance, but it omits several beneficial functionalities such as interactive programming, automatic memory management, and high-level data abstractions.

However, an ideal VM should support all those functionalities since they are beneficial. For example, efficient execution performance by dynamic compilation enhances the variety of applications that can be developed for microcontrollers. Consider an application that performs signal processing triggered by a timer interrupt occurring every 0.1 seconds. The signal processing must be completed within a 0.1-second interrupt cycle. If the execution speed is slow, it may be feasible to use only simple algorithms such as moving average filters. In contrast, if the execution speed is fast, more advanced algorithms like fast Fourier transforms (FFT) or finite impulse response (FIR) filters may become feasible. Furthermore, it may be possible to execute multiple filters concurrently.

Although run-time execution performance is particularly critical when programming microcontrollers, other VM functionalities are also important. For example, interactive programming enables programs to be written incrementally through an interface like Jupyter Notebook [21] or browser consoles. Popular interactive-programming environments for microcontrollers are REPLs (Read-Eval-Print Loops) such as MicroPython's and Espruino's [22]. Those interactive environments allow users to write programs more efficiently through a process of trial and error. This is particularly useful when programming microcontrollers, which often involves interacting with peripheral hardware devices. This fact makes rapid and iterative modifications especially beneficial. Proportional-Integral-Derivative (PID) control, which is frequently used in robot control, is a good example of leveraging the benefits of an interactive environment. PID control is a type of feedback control that calculates the control input based on the error between a target value and the current value, and it is used to accurately control the speed or position of a robot. For instance, it can be used to accurately stop a robot at a designated line. The control output in PID control is determined by the sum of the error (Proportional term, $P$), the integral of the error (Integral term, $I$), and the rate of change of the error (Derivative term, $D$), each multiplied by respective tuning parameters. If the tuning parameters are not set correctly, the robot may overshoot the target position, oscillate, or fail to move. Since these parameters depend on the characteristics of the controlled system, such as the motor's response, they must be experimentally determined through trial-and-error using the actual device. Traditional development environments using the C language are non-interactive programming environments. In these environments, every parameter adjustment requires recompiling a program and writing it to the flash memory of the microcontroller. They are significantly time-consuming. Conversely, an interactive environment allows programmers to immediately run a program and verify that code changes, such as updates to constant values, behave as expected while they are editing the program. This enables efficient parameter tuning.

Another beneficial VM functionality is, for example, language support for advanced type inference so that programs can be safely executed without a large number of programmers' explicit type annotations. This would help rapid prototyping. Automatic memory management, or garbage collection, is another example of beneficial functionality. It would also help safe program execution.

To address the trade-off between VM size and functionality, offloading VM components to another machine has been a promising approach. The offloaded components can access the computing resources of the machine hosting the components and provide useful functionalities without the constraints of the limited processing power and memory of microcontrollers. Potential offloading destinations are server computers connected to microcontrollers via the internet. Other destinations are laptop PCs connected through a serial cable to a microcontroller when developing application programs. These machines typically have orders of magnitude more computing resources than a microcontroller.

However, existing research activities and VMs have only offloaded only limited kinds of components. A dynamic or JIT compiler is a popular candidate for offloading. A number of VMs offload a dynamic compiler to servers connected via the internet [3, 4, 6, 8, 12, 19, 27]. The other components of those VMs, however, remain on a microcontroller. For other components, a Wasm VM designed for microcontrollers, WARDuino [9], offloads a debugging component to another machine [13]. A Ruby implementation for microcontrollers, mruby [26], offloads a parser. We find several attempts to offload a VM component chosen in an ad hoc manner based on immediate needs, but no systematic approaches. Since the reduction in VM size by those attempts is limited, a majority of industrial programming environments for microcontrollers prioritize execution speed over interactive programming, and they are still non-interactive ones based on cross compilation by the C/C++ compiler.

A live programming system MμSE [16] offloads nearly all VM components to a host machine to support exploratory programming. In this system, a program is executed on the host machine while only sensors and actuators run on a microcontroller. A drawback of this system is that it does not allow microcontrollers to operate independently of the host machines, as programs are not executed directly on the microcontrollers. Ideally, a VM should be able to operate solely with the components available on the microcontrollers once the program development is complete and the microcontrollers are detached from the host machines.

## 3   BlueScript

We propose a disaggregated VM that offloads most of the VM components to a host machine through a wireless network but remains operational even after being disconnected from the host machine. Here, a host machine is a computer with a faster processor and larger memory than a microcontroller, typically, a laptop PC. On a microcontroller, the disaggregated VM retains minimum components that are necessary for executing a program. It offloads the remaining resource-consuming components to a host machine, and it thereby provides rich features such as efficient run-time execution performance and interactive programming even on the memory-contrained microcontrollers.

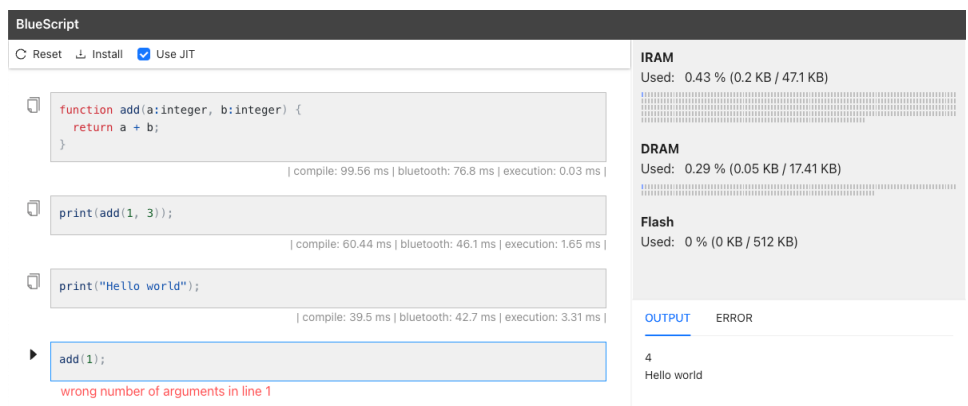

**Figure 1** An interactive shell for BlueScript.

To demonstrate the concept of the disaggregated VM, we present the BlueScript VM. BlueScript is our TypeScript-like small language for microcontrollers. The BlueScript VM provides a Read-Eval-Print Loop (REPL) for interactive programming, but it is not a traditional interpreter. To make more VM components offloadable, the REPL reads a code fragment through an interactive shell, compiles it by an incremental compiler, links the resulting binary code with the existing codebase, executes it, and prints the output. All of these components can be offloaded except for the run-time routines for executing the binary code. Furthermore, the REPL periodically runs a dynamic compiler to recompile frequently executed functions to be more efficient binary code. The dynamic compiler is also offloadable, whereas a run-time profiler is not as it must locally count the frequency of function invocations on the microcontroller.

A crucial issue for implementing disaggregated VMs is communication overhead among distributed components. Offloading VM components reduces a memory footprint on the microcontroller, but it involves communication costs between the microcontroller and the host machine. To mitigate this overhead, the BlueScript VM introduces a new VM component named *a shadow machine*, which enables efficient communication over the Bluetooth wireless network. The shadow machine is a virtual machine image shared between the microcontroller and the host machine. It exists on the host machine and reflects the current execution state of the whole disaggregated VM, including the state on the microcontroller. Whenever a VM component operates, the resulting state changes are reflected in the shadow machine. Other VM components read these changes from the shadow machine. This unification of the data source for the VM states reduces communication overheads by incrementally transferring newly generated differences and utilizing idle time for data transmission.

## 3.1 BlueScript

BlueScript is a TypeScript-like language that provides an interactive programming environment for microcontrollers. While BlueScript borrows its syntax from TypeScript, its semantics are much closer to more *static* languages such as Java so that a BlueScript

■ **Listing 1**  A BlueScript program for blinking an LED

```
1  import { GPIO } from 'gpio'
2  import { setInterval } from 'timer'
3
4  let led: GPIO = new GPIO(23);
5  let isLit: boolean = false;
6
7  function toggleLED() {
8    if (isLit)
9      led.on();
10   else
11     led.off();
12   isLit = !isLit;
13 }
14
15 setInterval(toggleLED, 1000);
```

program can efficiently run on a microcontroller. BlueScript is a managed language. An unused object is automatically reclaimed by a garbage collector.

An interactive shell for BlueScript has a notebook-like interface on a web browser. BlueScript programmers can incrementally write and execute code through this interface. Figure 1 illustrates this user interface of the interactive shell. Programmers write code in a cell on the left side of the screen and execute the code written in the cell by clicking the execution (triangle) button on the left side of the cell or by pressing Shift+Enter. The result of the execution is displayed in the OUTPUT and ERROR areas in the lower right corner of the screen. The memory usage on a microcontroller is displayed in the upper right corner of the screen. This interactive shell allows programmers to incrementally develop a program through an iterative process of executing program fragments, observing the results, and subsequently modifying the fragments. This is particularly useful when programmers are developing a program for controlling a small device or robot and they need to refine the program while observing the behavior of the controlled device or robot.

BlueScript supports a small but basic subset of TypeScript's syntax. It supports function, class, let, const, if-else, => (arrow functions), and so on, but it does not support outdated syntactic constructs such as var and with. Listing 1 shows an example of BlueScript program. It can be read mostly as if it is a normal TypeScript program, but the semantics of BlueScript is more *static* than TypeScript. Most dynamic language features available in TypeScript are unsupported in BlueScript. For example, BlueScript does not support eval, apply, or prototype-based inheritance.

The primitive data types in BlueScript are integer (30-bit signed integer), float (30-bit floating-point number), string (character string), boolean, and undefined. In BlueScript, number is an alias for integer, and null is an alias for undefined. They are identical, respectively. The language also supports arrays and class instances. The class system is static with single inheritance only and no interface support. An object must be created by the new operator as an instance of a class. An object literal such as { x: 3, y: 4 } is not supported. A method is not an object's property as a method is not a field in Java. All the instances of the same class share the same set of methods. A class definition may not be modified during run time, although a function definition is redefinable. The language does not allow adding a new method to an existing class or adding a new property to an existing object.

In BlueScript, all variables and functions must be statically typed. The language does not allow unsound typing. It does not support structural typing. A class type is nominally treated, and the subtype relationship is derived solely from type names. Interface types have not been supported yet. The design goal of BlueScript is to provide a simple static object system similar to those of C++ and Java so that method dispatch can be implemented through the classic use of virtual function tables. When a property is accessed, the relative location of that property within an object is statically determined, and thus no property name lookup is required at run time. When an array element is accessed, however, run-time range checking is performed on an array index. The type of a value assigned to an array element is not dynamically checked because array types are invariant. An array type T[] is not a subtype of S[] even when T is a subtype of S.

The language employs a simple form of gradual typing, inspired by work by Siek and Taha [24]. Gradual typing enables the mixing of static and dynamic typing in a single program. The programmer controls whether the type checking is static or dynamic by omitting or providing type annotations. In BlueScript, dynamically typed values are of the type named any, which is consistent with any other type. BlueScript performs type inference when type annotations are missing, inferring the type any if no other types can be determined. While the BlueScript VM skips run-time type checking for statically typed parts of a program, it performs type checking at run time for dynamically typed parts, resulting in performance overhead.

Since reducing the performance overhead incurred by run-time type checking is an aim of dynamic compilation mentioned below, we present an overview of the implementation of the types in BlueScript. A value of the type any, which is a value in the dynamically typed parts, is implemented by a technique called *pointer tagging*. It is a 32bit value where the lower 2 bits are used as a type tag. When a value is passed across the boundary between a dynamically typed part and a statically typed part, that is, when the type of a value is converted between the type any and the other types, the value is converted between a tagged value and an untagged raw value. Thus, a value of the integer type is treated as a raw 32bit integer value (the upper 2bits will be lost when the value is converted into a value of any), and a value of the float type is treated as a raw 32bit floating-point number (the 8bit exponent will lose 2bits by conversion). Arithmetic computing on those values does not imply run-time overheads due to type checking. A value of object types, such as string, classes, arrays, and functions, is represented by a tagged value as a value of any is. It is checked at run time whether the type of the object value is a subtype of a concrete type $T$ when it is converted from any to $T$. This run-time overhead is not large since BlueScript adopts nominal typing. Every object contains a pointer to the data structure representing its type, and run-time type checking only compares this pointer in most cases.

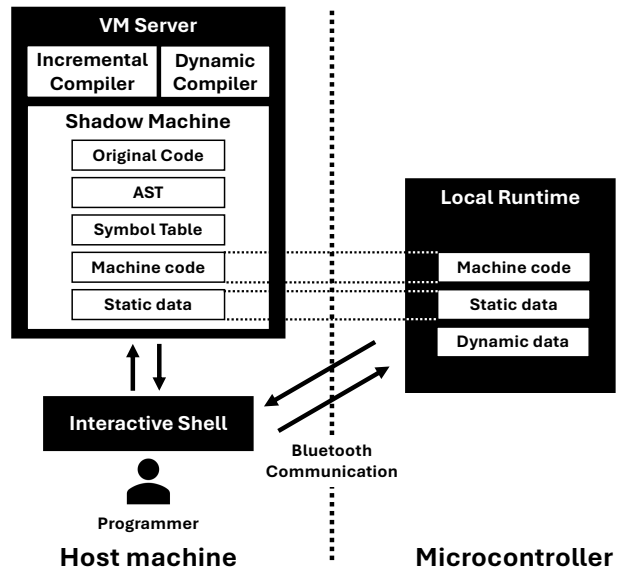

■ **Figure 2**  The architecture of the BlueScript VM

BlueScript supports the import declaration to dynamically import functions, classes, and global variables, which are exported by another module. Programmers can use import for dynamically loading library code. BlueScript currently provides several hardware libraries such as timer, gpio, button, and display. When such a library is loaded, the VM loads not only a BlueScript program but also the native-code library on which the BlueScript program depends.

### 3.2 Shadow Machine

As shown in Figure 2, the BlueScript VM consists of three sub-systems: an interactive shell, a VM server, and a local runtime. The former two sub-systems run on a host machine, while only the last one runs on a microcontroller. The interactive shell, implemented as a React application on a web browser, allows programmers to interactively develop a program. The VM server hosts an incremental compiler and dynamic compiler offloaded from the microcontroller. The local runtime executes compiled binary code while performing several background tasks such as garbage collection and profiling.

The interactive shell is also a networking hub for the three sub-systems. The local runtime communicates with only the interactive shell through Bluetooth, but it does not directly communicates with the VM server. It communicates with the VM server via the interactive shell. The interactive shell uses the Web Bluetooth API when communicating with the local runtime while it uses the HTTP with the VM server, which is a node.js application.

The VM server maintains a VM component named *a shadow machine* to mitigate communication overheads between the VM server and the local runtime. The shadow machine is a data structure representing the current execution states of the BlueScript

VM. It is shared among the VM components in the VM server and the local runtime. The execution states include the original source code, the abstract syntax trees, the symbol table, the compiled native code, memory layout, and profiling data. Several states such as the compiled native code are synchronized and reflected in the microcontroller to keep consistency.

When the interactive shell is given a code fragment, it sends the code fragment to the shadow machine in the VM server. The incremental compiler running in the VM server reads the code fragment from the shadow machine, parses it, and compiles it into native code. Then, a linker offloaded into the VM server resolves jump addresses, offsets, and similar references in the native code, by referring to the symbol table, which is also in the shadow machine. The linker stores the resulting binary code in the shadow machine. This binary code is sent from the shadow machine to the local runtime on the microcontroller as it is a state change to the code region of the BlueScript VM. It is reflected on the physical memory on the microcontroller. Neither the incremental compiler nor the linker directly writes to the microcontroller's memory. The dynamic compiler also reads and writes to the shadow machine to recompile frequently invoked functions. The recompiled functions are selected based on profiling data. The profiler runs in the local runtime and counts how frequently each function is invoked in BlueScript. The resulting counts are sent to the shadow machine, and they are used by the dynamic compiler. Note that the shadow machine currently does not maintain the execution state representing the heap memory on the microcontroller because it is not necessary for the incremental compiler or the dynamic compiler. Thus, the resulting state changes by garbage collection are not synchronized or reflected in the shadow machine.

To reduce the overhead due to the synchronization between the shadow machine and the local runtime on the microcontroller, the data transfer is performed asynchronously. State changes are not immediately transferred to/from the microcontroller for the synchronization. They are accumulated and transferred at once when the microcontroller enters an idle sate to await hardware interrupts or events, such as a timer interrupt. To prevent race conditions, when the VM components require modifying the data in the shadow machine, for example, when they modify the existing native machine code, they are restricted to only appending new native code to the code region in the shadow machine. Due to this restriction, the appended code includes a small machine-code routine that actually modifies the data when it is invoked. The local runtime invokes this routine when it enters an idle state in which the native code on the microcontroller can be safely modified. When the user redefines a BlueScript function through the interactive shell, the new function body is compiled and its native code is appended to the code region. The BlueScript VM switches from the original body to the new one by updating a global variable that points to the function body. This update is performed by the routine included in the appended code, which is invoked by the local runtime.

### 3.3 Offloading an Incremental Compiler

We adopt an incremental compiler [10] for implementing the BlueScript REPL. We deliberately avoid adopting a conventional interpreter-based approach, which does not generate native machine code. A conventional interpreter cannot be offloaded from the microcontroller. If it were offloaded, only the sensors and actuators would remain on the microcontroller, which would not run as a standalone device; it would require a constant connection to the host machine. This does not align with our objective.

The incremental compiler is different from ahead-of-time (AOT) compilers and dynamic (or just-in-time, JIT) compilers. A difference from AOT compilers is that AOT compilers compile entire programs *before* their execution whereas the incremental compiler incrementally compiles code fragments given *after* program execution starts. A difference from typical dynamic compilers is that dynamic compilers recompile existing code for improving its execution performance when it is frequently executed. Jikes RVM [1] does not include a bytecode interpreter, but it relies solely on two dynamic compilers for execution: the baseline compiler and the optimizing compiler. Its baseline compiler can be regarded as a variant of the incremental compiler although the incremental compiler for the BlueScript VM directly does not compile bytecode but source code.

The incremental compiler is offloaded into the VM server on the host machine. When the execution button is clicked (see Figure 1), a program text written in a cell of the notebook-like interface, which is usually one or a few lines, is sent through the shadow machine from the interactive shell to the incremental compiler in the VM server. It is parsed, type-checked, and compiled into native machine code by the incremental compiler. Then, the native code is linked to the previously generated native code. All the operations are performed on the host machine through a shadow machine. After linking, the generated native machine code is transferred to the local runtime on the microcontroller. Its execution kernel solely stores the received native code in memory and jumps to the entry point of that code for execution. Only the execution of native machine code is performed on the microcontroller while the other operations are performed on the host machine. Note that the local runtime does not perform linking. It simply stores the code in memory *as it is*. If the code prints a message or throws a run-time error during the execution on the microcontroller, that output is sent back to the interactive shell. It is displayed in the OUTPUT or ERROR section of the interactive shell (See Figure 1).

The incremental compiler is implemented in TypeScript. For compilation, it exploits an existing compilation toolchain. After parsing and type checking, it stores the resulting abstract syntax tree in the shadow machine, and translates a BlueScript program into a program in the C language. Then, the C program is compiled into native machine code by the off-the-shelf cross-compiler for the target microcontroller. This design decision is chosen to reduce our engineering efforts and facilitate portability when migrating to new microcontroller architectures. The resulting native code is stored in the shadow machine with its symbol table information, which includes the names, types, and addresses of functions and variables in the native code. The memory address of the native code is determined by the incremental compiler. Then, the symbol table information is analyzed to identify external references, and a linker script is generated. The off-the-shelf linker is executed with the linker script for resolving references and symbols in the native code. The result of linking is also reflected on the symbol table information in the shadow machine.

The BlueScript VM provides several libraries for accessing the microcontroller's hardware. They are BlueScript programs but depend on other C libraries. These C libraries are precompiled on the host machine. When a BlueScript program executes the import declaration, the native machine code for that BlueScript program is linked with the native machine code for the imported BlueScript library and also the C library on which the BlueScript library depends. Then, all the native code is sent to the local runtime running on the microcontroller except for the code that has already been sent. This on-demand code transferring is useful for reducing memory consumption on the microcontroller since the code size of some C libraries for accessing hardware such as a display and Wi-Fi is significantly large.

Although the BlueScript VM is a disaggregated VM, it can run solely with the local runtime running on the microcontroller. The microcontroller can be detached from the host machine, for example, after program development completes. The BlueScript VM is reduced to include only the components that are not offloaded, and continues to run. It no longer accepts new code fragments, but the existing native machine code is saved into the microcontroller's flash memory. This saved code is executed repeatedly each time the microcontroller restarts. To detach the microcontroller from the host machine, the programmer clicks the install button on the web page for the notebook-like interface. Then the interactive shell concatenates all the code written in the cells in the web page, and it sends the concatenated code through the shadow machine to the incremental compiler running in the VM server. The incremental compiler compiles the concatenated code, and it links the resulting native machine code with the libraries and the code for the local runtime. The native code after linking is stored back in the shadow machine, and then it is transferred to the local runtime on the microcontroller. The local runtime saves the received native code into the flash memory for the microcontroller. Then the microcontroller is rebooted and the saved code is executed. Note that the local runtime usually save the received native code into SRAM.

### 3.4 Offloading Dynamic Compilation

The BlueScript VM offloads a dynamic compiler, or a Just-In-Time (JIT) compiler. Only the VM component for collecting profiling data remains on the microcontroller. The collected profiling data are sent to a dynamic compiler in the VM server on the host machine through the shadow machine, and the dynamic compiler generates optimized code and sends it back to the microcontroller.

The current dynamic compiler of the BlueScript VM optimizes functions that receive parameters of the type any, which represents their types are not statically known. It specializes in their function bodies to receive parameters of specific concrete types so that overheads due to type conversion will be eliminated. The profiler on the microcontroller counts the number of calls to such functions, which are statically selected by the incremental compiler. Suppose that the following add function is a candidate of specialization:

```
1  function add(a, b) { return a + b }
```

The parameter types for a and b are any. This function is compiled into the native code corresponding to this C code:

```
1  static value_t original_fbody_add(value_t self, value_t _a, value_t _b) { ... }
2  struct func_body original_add = { original_fbody_add, "(aa)a" };
3  static value_t fbody_add(value_t self, value_t _a, value_t _b) {
4      static uint8_t call_count = 0;
5      static uint32_t type_count[NR_ARGS + 1][N] = {};
6      PROFILE(0, &call_count, &type_count, _a, _b, VALUE_UNDEF, VALUE_UNDEF);
7      return ((value_t (*)(value_t, value_t, value_t))original_add.fptr)(self, (_a), (_b));
8  }
9  struct func_body _add = { fbody_add, "(aa)a" }
```

original_fbody_add in line 1 implements a function body returning a + b. value_t in the C code represents the type any in BlueScript. Line 4 to 6 are for profiling. When the add function is called, a function pointer to fbody_add is obtained from _add in line 12 for call. When the calling count exceeds the threshold (5 in the current implementation), the profiler starts collecting the concrete argument types passed to the function. It checks the 2-bit tags of the passed values to obtain their actual types. Once the type information is sufficiently collected, the most frequently passed group of argument types is sent to the dynamic compiler as the target types for specializing the body of the function.

Then the dynamic compiler generates a specialized body of the function to receive the parameters of the given group of concrete types. This generation is performed using the shadow machine, which holds an abstract syntax tree for the function body. The return type of the specialized body does not change. It is the same as the original type. The guard code is also generated. It performs dynamic dispatch to an appropriate function body; if the arguments match the given group of concrete types, the specialized body is invoked. Otherwise, the original body is invoked. For the example above, the following code is generated:

```
1   static value_t specialized_fbody_add(value_t self, int32_t _a, int32_t _b) { ... }
2   struct func_body specialized_add = { specialized_fbody_add, "(ii)a" };
3   static value_t fbody_add(value_t self, value_t _a, value_t _b) {
4       if (is_int_value(_a) && is_int_value(_b))
5           return ((value_t (*)(value_t, int32_t, int32_t))specialized_add.fptr)(self,
6                                       safe_value_to_int(_a), safe_value_to_int(_b));
7       else
8           return ((value_t (*)(value_t, value_t, value_t))original_add.fptr)(self, (_a), (_b));
9   }
10  void bluescript_main4_() { ... _add.fptr = fbody_add; }
```

■ **Table 1** Benchmarks programs (the upper programs are from "Are we fast yet?" and the lower ones are from ProgLangComp)

| Name | LoC | Description |
|------|-----|-------------|
| Bounce | 93 | Simulates a ball bouncing within a box. |
| List | 79 | Recursively creates and traverses lists. |
| Mandelbrot | 83 | Calculates the classic fractal. |
| NBody | 185 | Simulates the movement of planets in the solar system. |
| Permute | 42 | Generates permutations of an array. |
| Queens | 61 | Solves the eight queens problem. |
| Sieve | 38 | Finds prime numbers based on the sieve of Eratosthenes. |
| Storage | 57 | Creates and verifies a tree of arrays to stress the garbage collector. |
| Towers | 83 | Solves the Towers of Hanoi game. |
| CD | 1053 | Simulates an airplane collision detector. |
| Richards | 523 | Simulates an operating system kernel. |
| CRC | 113 | Computes CRC32-IEEE checksums using a precomputed table. |
| FFT | 97 | Computes FFT (fixed point, complex pair of int16). |
| FIR | 55 | Converts waveform (floating point) by an FIR filter. |
| IIR | 43 | Converts waveform (floating point) by a digital biquad filter. |
| SHA256 | 175 | Computes SHA256 hashes. |

We assume that the profiler reports that the `add` function is usually called with two integer values for `a` and `b`. `specialized_fbody_add` in line 1 implements a specialized function body receiving two integer values. New `fbody_add` in line 3 is the guard code.

The generated code is compiled and linked by the incremental compiler, and it is sent back through the shadow machine to the microcontroller, where the local runtime substitutes the new code for the original code. For the example above, `bluescript_main4_` in line 10 performs this substitution by updating the function pointer in `_add`. The local runtime calls it when the microcontroller enters an idle state.

When the programmers redefine a function that has been already specialized by the dynamic compiler, the new function body is simultaneously specialized with the same profiling data. Because the caller sites do not change, the new function body is anticipated to be as frequently called with arguments of the same types as the original one.

## 4 Experiments

We conduct experiments to assess whether the benefits of an incremental compiler and a dynamic compiler are retained despite the potential overhead of offloading. The primary benefit of the incremental compiler is to achieve faster execution speeds than interpreter-based languages while maintaining interactivity. The benefit of the dynamic compiler is its ability to improve performance during run time.

We use an IoT development kit, M5Stack Fire, based on the ESP32-DoWDQ6 microcontroller with 520 KB of RAM, 16 MB of Flash memory, and 8 MB of PSRAM. For a host machine, we use MacBook Pro, which is equipped with Apple M1 Pro, 16 GB of memory, and 512 GB of storage. The off-the-shelf compiler used by the incremental compiler is the xtensa-esp32-elf-gcc (crosstool-NG esp-2022r1) 11.2.0 compiler with the -O2 option.

### 4.1 Execution Time

We first conduct a performance experiment to assess whether BlueScript's execution speed is faster than that of widely used MicroPython and Espruino, even with the incremental compiler offloaded. We compare the BlueScript VM with MicroPython V1.19.1 and Espruino 2V20. Espruino is a virtual machine for JavaScript. We also compare with the C++ compiler used as the off-the-shelf compiler for the BlueScript VM. We use two benchmark suites "Are We Fast Yet?" [15][2] and ProgLangComp [20].[3] Each benchmark programs are listed in Table 1. In "Are We Fast Yet?", we use all the micro benchmarks and only the two macro benchmarks due to lack of manpower for porting. ProgLangComp is a collection of MicroPython programs for evaluating the performance of the ESP32 microcontroller. They are typical application programs for microcontrollers such as signal processing and hash functions. In ProgLangComp, we use input of length 1024. We port these programs into BlueScript. For BlueScript, we write two programs for each benchmark program. We call them *a typed version* and *an untyped version*. In typed versions, all functions and variables are given type annotations. In untyped versions, we remove type annotations from all the function parameters while retaining type annotations for variables and properties. We do not write untyped versions for CRC, FFT, or SHA from the ProgLangComp benchmark suite because they assume that integers are 32 bits. We write typed versions for them because the current BlueScript VM deals with an integer value as a 32-bit value unless that value is converted into a value of the type any.

To run a BlueScript program, we write the entire program in a single cell of the notebook-like interface and click the execution button so that the entire program will be simultaneously compiled together. We turn off the dynamic compiler for this experiment. For MicroPython and Espruino, we run benchmark programs according to the documents on their official pages. We download their VMs from the official page and install them on the microcontroller before rebooting. Then we write each benchmark program on the host machine and send it to the microcontroller for execution. We use the ESP-IDF build tool for the C++ language to compile and execute benchmark programs. We give the -O2 option to the compiler.

Figure 3 and Figure 4 show the results. The execution of CD and Richards by Espruino are timed out. All the execution times are the means of five runs. Bars represent the mean of five measurements, and error bars indicate the standard error of the mean. Note that the Y axes are log scales. The results show that even with the incremental compiler offloaded, BlueScript execution speed is even faster than, MicroPython and Espruino. BlueScript is one to two orders of magnitude faster than MicroPython and three to four orders of magnitude faster than Espruino. The performance difference

---

[2] https://github.com/ignasp/ProgLangComp_onESP32 (Visited on 2025-09-29).
[3] https://github.com/smarr/are-we-fast-yet (Visited on 2025-09-29).

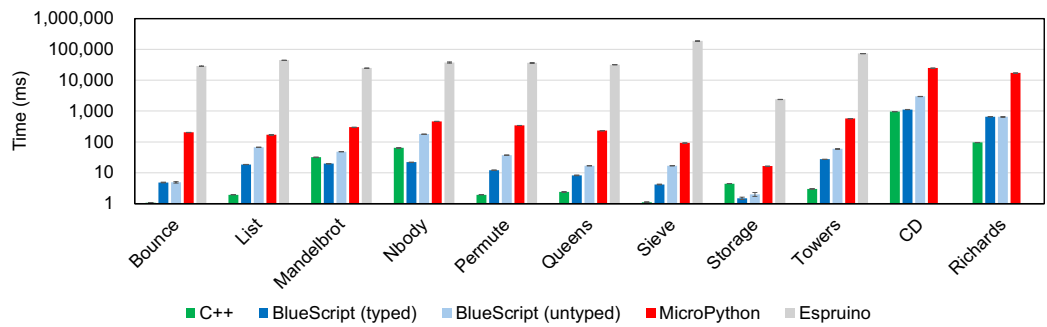

**■ Figure 3** Execution time of "Are we fast yet?" benchmark suit

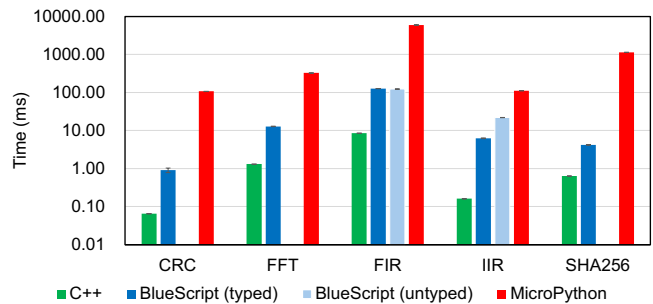

**■ Figure 4** Execution time of `ProgLangComp` benchmark suit

is much larger when a program involves a large number of arithmetic operations on primitive types such as Bounce, FIR, CRC, and SHA256. This result is because the BlueScript programs are type-annotated and thus they less frequently perform run-time type checking and conversion (in the untyped version of Bounce, function parameters are not type-annotated, but variables are still type-annotated). Compared to the C++ language, BlueScript is up to 10 times slower. The slowdown is significant when a program involves a large number of function calls since function calls in BlueScript involves indirect memory accesses. Such programs are Towers and List. On the other hand, for benchmarks like Mandelbrot, NBody, and Storage, BlueScript outperforms C++. For Mandelbrot and NBody, this is likely because the C++ versions use double for calculations, whereas BlueScript employs int32_t integers. Furthermore, in the C++ Storage benchmark, performance is hindered by more time-consuming memory allocation, making it slower than BlueScript, which uses a simpler mark-and-sweep garbage collection. BlueScript (untyped) is several times slower than BlueScript (typed). The slow down in NBody is particularly noticeable; while the slow down in the other benchmarks is around three times slower, NBody is eight times slower. This is likely due to the frequent accesses to an statically-untyped object, which incur performance overheads from property-name lookups.

■ **Table 2** Benchmark programs for interactivity

| Scenario | frag-ments | LoC | Description |
|---|---|---|---|
| Dice | 4 | 23 | Display a random integer from 0 to 6 when button pressed. |
| Flashing Heart | 3 | 20 | Display a large heart icon and a small heart icon alternately. |
| Love Meter | 3 | 22 | Display "Love meter" and a random integer when button pressed. |
| Name Tag | 4 | 13 | Display a programmer's name. |
| Smiley Buttons | 3 | 17 | Display happy on button B press, and sad face on button C press. |

■ **Listing 2** The Smiley Buttons program in BlueScript

```
1  // Code Fragment #1
2  import { Display } from 'display';
3  import { buttonOnPressed } from 'button'
4
5  const BUTTON_PIN_B = 38;
6  buttonOnPressed(BUTTON_PIN_B, () => {});
7  print("$$");
8
9  // Code Fragment #2
10 let dspl = new Display();
11 const colorWhite = dspl.color(255, 255, 255);
12 const colorRed = dspl.color(255, 0, 0);
13 buttonOnPressed(BUTTON_PIN_B, () => {
14   dspl.showIcon(dspl.ICON_HAPPY_FACE, colorRed, colorWhite);
15 });
16 print("$$");
17
18 // Code Fragment #3
19 const BUTTON_PIN_C = 37;
20 buttonOnPressed(BUTTON_PIN_C, () => {
21   dspl.showIcon(dspl.ICON_SAD_FACE, colorRed, colorWhite);
22 });
23 print("$$")
```

## 4.2 Interactivity

To assess the impact of the offloaded incremental compiler on the interactivity of BlueScript compared to MicroPython, we compare the response time of BlueScript's interactive shell and MicroPython's Wi-Fi-based REPL. The latter is a browser-based interface to remotely control the MicroPython VM through Wi-Fi although BlueScript does through Bluetooth. We incrementally write a program step by step according to a given scenario, and measure the response time for every step. We take five scenarios

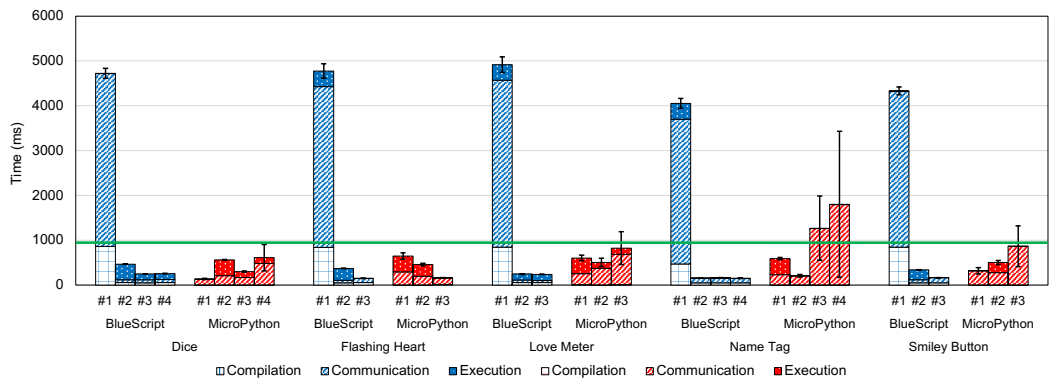

■ **Figure 5**  Response time for every code fragment from a tutorial course on the website of `MakeCode`,[4] which provides programming lessons using the `micro:bit` [7] microcontroller for beginners. We write two versions of programs for each scenario in BlueScript and MicroPython. Table 2 lists five scenarios. It also presents the number of lines for each BlueScript programs.

Each scenario consists of three or four code fragments. In each scenario, a programmer repeatedly writes a code fragment and runs it to test its results. Listing 2 shows code fragments in BlueScript for the scenario called Smiley Buttons. This scenario consists of three code fragments. A programmer types and runs each code fragment step by step to finally build a small application program that shows happy and sad faces on the screen. The code fragment #1 imports display and button libraries and binds an empty function to button B. The code fragment #2 initializes the display module and binds a function for displaying a happy face icon to button B. The code fragment #3 binds a function for displaying a sad face icon to button C. Every code finally prints $$ to notify the end of the execution to the host machine.

For BlueScript, we write and execute code fragments one by one in cells in the interactive shell. For MicroPython, the code in each code fragment is transformed into a single line since the REPL of MicroPython receives only a single line at once. We combine all the statements in the code fragment into a single line where statements are separated by a semicolon. Since all the five scenarios finally display an icon and text on the screen, we implement BlueScript and MicroPython libraries to manipulate the screen and the buttons of our target device `M5Stack Fire`. These libraries call the common C library that we implement through a similar interface.

We execute every code fragment seven times and measure the response time by measuring the time elapsed from pressing the execution button until $$ is displayed on the screen. Note that every code fragment prints $$ at the end as shown in Listing 2. We also measure the execution time and the compilation time. We average the response time, compilation time, and execution time from these seven runs. We also calculate the standard error of the response time. We then calculate the communication time by subtracting the compilation and execution times from the response time.

---

[4] https://makecode.microbit.org/ (Visited on 2025-09-29).

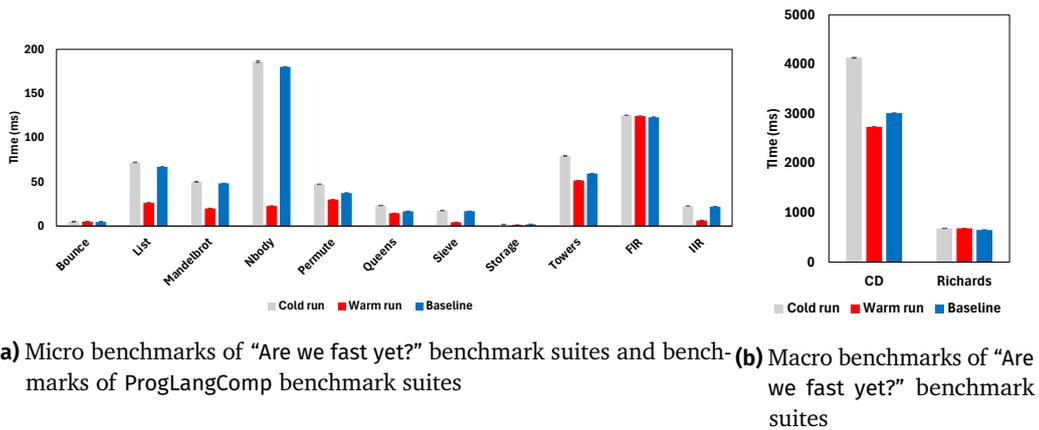

**(a)** Micro benchmarks of "Are we fast yet?" benchmark suites and benchmarks of `ProgLangComp` benchmark suites

**(b)** Macro benchmarks of "Are we fast yet?" benchmark suites

■ **Figure 6**  Execution times of the cold run, warm run and baseline

Figure 5 shows the response time for every code fragment. The error bar shows the standard error of the response time. The green line represents 1 second. According to the literature [23], the users perceive the response as being immediate when it takes less than 0.5 to 1 seconds. The results show that the response time of BlueScript VM is under 550 msec. and the VM achieves as good responsibility as to MicroPython except for the first code fragment. The increase of the response time for the first code is because BlueScript must import libraries such as `Display`, `Button`, and `Timer` in the first cell. In particular, the size of the `Display` library used by all the benchmarks is large and hence transferring that library through Bluetooth causes longer response time. The code fragments #2 of Dice, #1 of Name Tag, and #1 of Smiley Button cause longer response time. The response time of MicroPython REPL is unstable, and we occasionally observe the response time suddenly drops. The cause of the drop down is not well understood, so we are still investigating.

### 4.3 Dynamic Compilation

To verify that the dynamic compiler improves execution speed even while being offloaded, we measure the performance changes when applying the dynamic compiler of the BlueScript VM. The following three cases are measured.

- **Cold run:** the first run with dynamic compilation enabled. This involves profiling overhead.
- **Warm run:** a run after 15 runs with dynamic compilation enabled.
- **Baseline:** a run with dynamic compilation disabled.

For the measurement, we use the same set of benchmark programs as in Section 4.1. For BlueScript, we only use the untyped versions of programs since the dynamic compilation does not work for the typed versions.

Figure 6 shows the execution times for each benchmark, averaged over five runs. Bars represent the mean of five measurements, and error bars indicate the standard error of the mean. By comparing the baseline and warm run, it is revealed that the dynamic compilation improves the execution performance up to 7.8 times. In

**BlueScript: A Disaggregated Virtual Machine for Microcontrollers**

■ **Listing 3**   The program for periodically measuring the elapsed time

```
1  import {setInterval, getTimeMs} from 'timer'
2
3  function benchmark() { ... omit ...}
4
5  const timerId = setInterval(()=>{
6     const start = getTimeMs();
7     benchmark();
8     const end = getTimeMs();
9     print(end - start);
10 }, 2000);
```

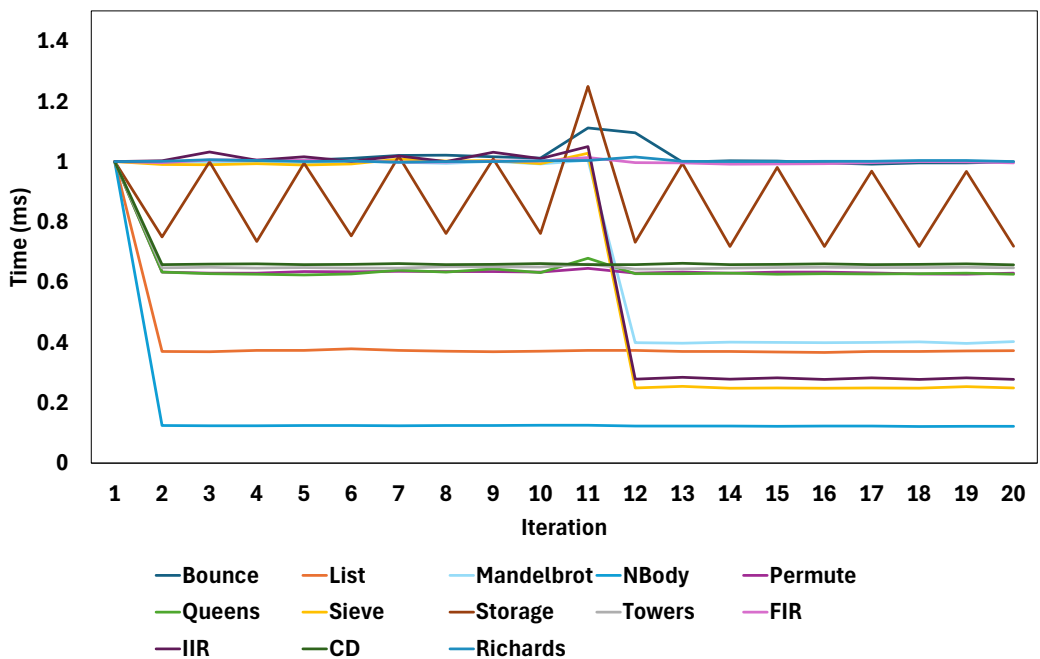

■ **Figure 7**   Latency of dynamic compilation. The values shown in the graph represent the execution time of each benchmark divided by its execution time in the first iteration.

particular, calculation-intensive benchmarks like NBody, Sieve, and IIR show significant performance improvements. This is consistent with the results in Section 4.1. Although Mandelbrot and FIR are also calculation-intensive, their kernel functions do rarely access function parameters. The existence of type annotations for function parameters has only little impact on execution performance. Bounce is also calculation-intensive, but its kernel function does not take a parameter. This program is not a target of the dynamic compilation.

We also measure the change of the execution time by dynamic compilation. The benchmark programs are the same as Figure 6. We use a program listed in Listing 3

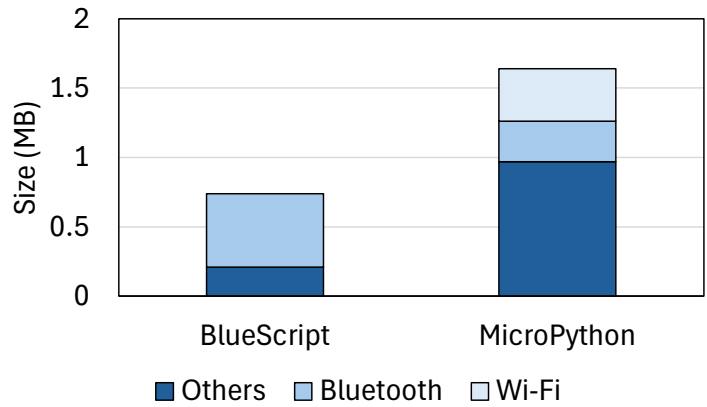

■ **Figure 8** The size of the run-time routines on a microcontroller.

to repeatedly run a benchmark program and measure its execution time every five seconds for macro benchmarks in "Are we fast yet?" benchmark suites (CD and Richards) and every one second for other benchmarks. Since a benchmark program is started on a timer interrupt, the microprocessor is idle between the executions of the benchmark program. The profiling data is transferred from the microcontroller during this idle time.

Figure 7 shows the results. Each execution time in the figure represents the average of five measurements. The values shown represent the execution time of each benchmark divided by its execution time in the first iteration. For Storage, the execution time is vibrating. This is because garbage collection tends to occur during execution roughly once every two runs. Since Storage has an inherently short execution time, the overhead from this garbage collection has a pronounced impact on the measured execution time, causing the observed vibration. For the six benchmark programs List, NBody, Permute, Queens, Towers, and CD, a marked reduction in execution time is observed between the first and second runs. This is because the functions specialized for optimization are frequently called and the call counts quickly exceed thresholds. For Mandelbrot, Sieve, and IIR, a reduction occurs around the 12th run. Since data transfer between the microcontroller and the host machine is scheduled to be executed between benchmark program's runs, the overheads due to the data transfer are not observable in Figure 7. In fact, if profiling data is immediately transferred when they are collected, we occasionally observe that the execution speed of benchmark programs slows down.

### 4.4 Memory Footprint

Figure 8 shows the memory footprint of run-time routines on a microcontroller. For BlueScript, their size is 0.74 MB, but it will be 0.27 MB if the Bluetooth library is excluded although the Bluetooth library is mandatory. For MicroPython, since the whole virtual machine resides on a microcontroller, the run-time routines constitute the entire machine. Their size is 1.64 MB, but it will be 1.40 MB if the Bluetooth library is excluded. It will be 1.26 MB if the Bluetooth library is included but the Wi-Fi library is excluded. The size of the virtual machine excluding both Bluetooth and Wi-Fi libraries is 0.97 MB. Note that these run-time routines are stored in not SRAM but flash memory, which is significantly larger than SRAM.

As seen in Figure 8, the memory footprint of the BlueScript VM on a microcontroller is smaller than MicroPython since components are offloaded. Unlike MicroPython, the BlueScript VM does not require initially including other hardware libraries than Bluetooth at the booting time. Programmers can dynamically load necessary hardware libraries on demand by import declarations as already mentioned in Section 3.1.

## 5  Related Work

Existing systems offload only a subset of VM components by modifying the original systems, or they attach a new VM component as an offloaded component to the original systems. To the best of our knowledge, there are no systems that are entirely designed using a novel architecture to offload VM components. Furthermore, existing systems offload only a limited kind of VM components.

Remote JITs [3, 4, 6, 12, 19, 27] and Espruino [8] offload JIT compilers from embedded and/or mobile devices to compilation servers connected via the internet. These JIT compilers compile all methods or frequently invoked methods but do not rely on run-time information such as type information. The BlueScript VM utilizes more run-time information, demonstrating the potential for more extensive dynamic compilation by an offloaded dynamic compiler for microcontrollers. Furthermore, those JIT compilers except for [12] do not offload linking, which is performed on the client-device side. Therefore, the information necessary for linking, such as the global offset table, should be sent from the server to the client device and the client device must maintain a symbol table. On the other hand, the BlueScript VM performs linking on the host machine using the shadow machine. The information for linking is not sent from a host machine to a microcontroller or stored in the microcontroller. Although VM⋆ [12] performs linking on the server side, the compilation server does not retain the source code. The client device must send bytecode to be compiled to the server with a compilation request.

Other VMs [11, 14] also offload dynamic compilation to other machines. They perform dynamic compilation based on profiling information gathered on the client devices. However, these VMs are not targeted at microcontrollers, and the experiments show that device-side memory consumption ranges from tens to hundreds of megabytes. Furthermore, since the server does not retain the source code, the client device must include the method's bytecode in a compilation request to be sent.

mruby [26] is a Ruby implementation for microcontrollers, and it offloads the parser. The mruby processor compiles the received source code into bytecode and then sends it back to the microcontroller for execution. mruby's VM component for interpreting intermediate code is retained on the microcontroller whereas the corresponding component in BlueScript is an incremental compiler and it is offloaded to the host machine. The approach for the BlueScript VM enables more powerful optimization during compilation and reduces memory consumption on the microcontroller.

EDWARD [13] serves as a remote debugger for WARDuino [9], a Wasm VM designed for microcontrollers. EDWARD offloads debugging capability to a host machine whereas the BlueScript VM offloads an incremental compiler and a dynamic compiler. The BlueScript VM does not have a debugger in the first place, so it is not offloaded to the host machine.

Static TypeScript [2] offloads an AOT compiler to the host machine whereas Blue-Script offloads an incremental compiler. Static TypeScript does not provide an interactive development environment. Furthermore, it does not include a dynamic (or JIT) compiler.

Sirer et al. also proposed a system called Distributed VM[25], which separates virtual machine components and places them on distinct network machines. However, in this system, the primary components separated from the client are the rule checking services (such as the bulk of the bytecode verifier and security policy enforcement mechanisms) and code transformation services (like the binary rewriter used for instrumenting code). Although the implementation of a dynamic (JIT) compiler was envisioned, the paper states it was unimplemented at the time of publication. Furthermore, the system, in its described form, does not provide features such as interactive development environments.

The work presented in this paper is an enhancement of the work previously presented in [18]. Although the incremental compilation for the BlueScript VM is being presented in that literature, it is a new contribution to show that the architecture of the BlueScript VM can be enhanced to support dynamic compilation. The idea of disaggregated VMs is also our new proposal. All the experiments shown in this paper are conducted with the latest version of the BlueScript VM.

## 6  Conclusion

We present an implementation of the BlueScript VM, a disaggregated VM offloading incremental compilation and dynamic compilation. In this implementation, a virtual machine component named *a shadow machine*, which mirrors the execution states on the microcontroller, helps efficient communication among offloaded and retained virtual-machine components. We conduct experiments and show that the execution performance of the BlueScript VM with offloaded incremental compiler is one to two orders of magnitude faster than that of MicroPython. Furthermore, its response time is maintained within 550 msec, showing that the BlueScript VM remains interactive with the offloaded incremental compiler. We also show that the offloaded dynamic compiler can speed up program execution by up to 7.8 times.

## Data-Availability Statement

The artifact associated with this paper is available at [17].

**Acknowledgements**   This work was supported by JSPS KAKENHI Grant Numbers JP20H00578 and JP24H00688, and JST SPRING Grant Number JPMJSP2108.

## About the authors

**Fumika Mochizuki** Contact her at fumika.maejima@csg.ci.i.u-tokyo.ac.jp.
https://orcid.org/0009-0005-2386-1369

**Tetsuro Yamazaki** Contact him at yamazaki@csg.ci.i.u-tokyo.ac.jp.
https://orcid.org/0000-0002-2065-5608

**Shigeru Chiba** Contact him at chiba@csg.ci.i.u-tokyo.ac.jp.
https://orcid.org/0000-0002-1058-5941